\documentclass{PoS}

\title{Matrix elements and diquark correlations in quenched QCD with 
overlap fermions.}

\ShortTitle{Matrix elements and diquark correlations in quenched 
QCD with overlap fermions.}

\author{Ronald~Babich$^a$, Nicolas~Garron$^b$,
        Christian~Hoelbling$^c$, Joseph~Howard$^a$, Laurent~Lellouch$^d$,
        and \speaker{Claudio~Rebbi}$\,^a$\\
        \llap{$^a$}Department of Physics, Boston University, Boston, MA\\
        \llap{$^b$}DESY, Platanenallee 6, 15738 Zeuthen, Germany\\
        \llap{$^c$}Department of Physics, Bergische Universit\"at 
                   Wuppertal, Germany\\
        \llap{$^d$}Centre de Physique Th\'eorique\thanks{UMR 6207 du CNRS 
                   et des universit\'es d'Aix-Marseille I, II et du Sud
                   Toulon-Var, affili\'ee \`a la FRUMAM.}, Marseille, France\\
        \\
        E-mail: \email{rebbi@bu.edu}}

\abstract{We present results for $B_K$ and selected matrix elements for
beyond the standard model interactions obtained in quenched QCD with 
overlap fermions.  We also illustrate results on baryon wave-functions 
and diquark correlations within baryons in the Coulomb and Landau gauge.}

\FullConference{XXIVth International Symposium on Lattice Field Theory\\
                July 23-28, 2006\\
                Tucson, Arizona, USA}

\renewcommand{\logo}{
\raisebox{-5mm}{
\parbox{\textwidth}{
\includegraphics{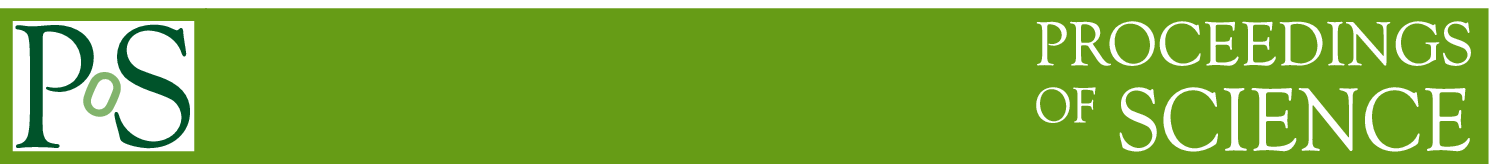}
\begin{flushright}
BUHEP-06-07 \\
CPT-P73-2006
\end{flushright} }}}

\begin{document}

\section{Introduction}

Over the last few years we have performed a study of quenched QCD with overlap 
quarks.  We generated 100 configurations of size $18^3\times 64$ at
$\beta=6$ and 100 configurations of size $14^3 \times 48$ at $\beta=5.85$,
with Wilson gauge action and separated from each other by 10,000 Metropolis
steps.  The corresponding lattice spacings, from the Sommer scale,
are $a^{-1}=2.12{\rm GeV}$ and $a^{-1}=1.61{\rm GeV}$.
These configurations were transformed to the Landau gauge,
and overlap quark propagators were calculated for a single point source
and all 12 color-spin combinations with $\rho =1.4$, 
$am_q=0.03, 0.04, 0.06, 0.08, 0.1, 0.25, 0.5, 0.75$ at $\beta=6$,
and $\rho =1.6$, $am_q=0.03, 0.04, 0.053, 0.08, 0.106,$ $ 0.132, 0.33, 
0.66, 0.99$ at $\beta=5.85$.  (We recall that the overlap
Dirac operator for a quark of mass $m$ is given by $[1-(am)/(2\rho)]D+m$,
where $D = (\rho/a)(1+\gamma_5 H(\rho) /\sqrt{H(\rho)^2})$ and
$H(\rho)$ stands for the Hermitian Wilson-Dirac operator
with mass $-\rho/a$.)  A Zolotarev approximation to the inverse square 
root was used for the first 55 configurations at $\beta=6$, then
a Chebyshev approximation of degree $100\sim 500$, after Ritz projection 
of the 12 ($18^3\times64$ lattice) or 40 ($14^3\times 48$ lattice)
lowest eigenvectors of $H^2$. The convergence criterion was  $|x^{-1/2}- 
\sum T_n(x)| <10^{-8}$, $\vert D^\dag D \psi -\chi\vert^2 <10^{-7}$.
The use of propagators with a point source, as opposed to extended
or wall sources, was dictated by our desire to calculate matrix elements
and renormalization factors, and by the limitation of the computational
resources at our disposal.  Results for light hadron spectroscopy
were presented at the Lattice 2005 Symposium~\cite{Babich:2005kt} and 
in Ref.~\cite{Babich:2005ay}.  We refer the reader to~\cite{Babich:2005ay} 
for details of our calculations and for references to other studies of 
lattice QCD with overlap quarks on large lattices.  We present
here results we recently obtained on $\Delta S=2$ matrix elements
and on baryon wave-functions and diquark correlations inside baryons.

\section{$\Delta S=2$ matrix elements}

We evaluated matrix elements of the operators
$ O_1=[\bar s^a\gamma_\mu(1-\gamma_5) d^a]
[\bar s^b\gamma^\mu(1-\gamma_5)d^b] ,
O_2=[\bar s^a(1-\gamma_5)d^a][\bar s^b(1-\gamma_5)d^b],
O_3=[\bar s^a(1-\gamma_5)d^b][\bar s^b(1-\gamma_5)d^a],
O_4=[\bar s^a(1-\gamma_5)d^a][\bar s^b(1+\gamma_5)d^b]$ and
$ O_5=[\bar s^a(1-\gamma_5)d^b][\bar s^b(1+\gamma_5)d^a]$
which are needed to study neutral kaon mixing in the standard model 
(SM) and beyond (BSM).  An extensive description of our work and 
results has been presented in~\cite{Babich:2006bh}.  Here we we 
only outline the main features of our analysis.  We would like 
to direct the reader to~\cite{Babich:2006bh} for a detailed 
bibliography of other relevant investigations.

\begin{figure}[h!]
\begin{center}
\includegraphics*[width=0.6\textwidth]{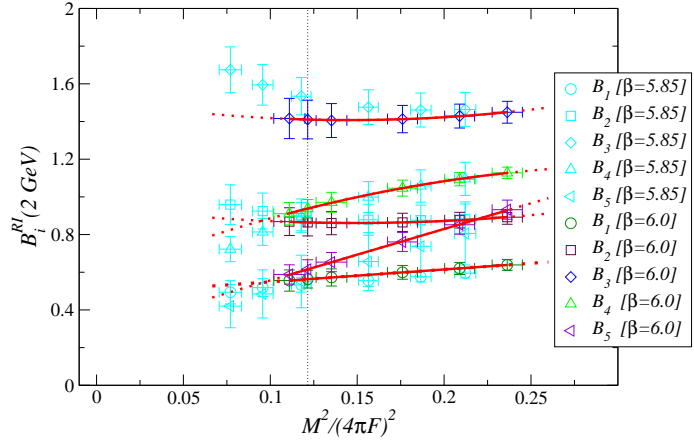}
\caption{\label{fig:biriinter600and585} Mass-dependence, in terms of
  the variable $M^2/(4\pi F)^2$, of the $B$-parameters $B_i$,
  $i=1,\cdots,5$, in the RI/MOM scheme at $2$~GeV. The solid curves
  are the results of the fits described in the text, and are plotted
  in the fit region. The fits are used to interpolate the results to
  the kaon point $M^2/(4\pi F)^2=M_K^2/(4\pi F_K)^2$, shown as a
  vertical dotted line.  The dashed curves are an extension of the fit
  curves outside the fit range.}
\end{center}
\end{figure}

\begin{figure}[h!]
\begin{center}
\includegraphics*[width=0.6\textwidth]{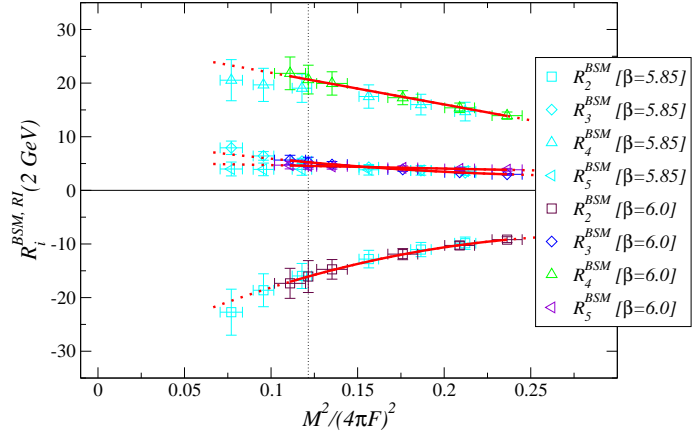}
\caption{\label{fig:riinter600and585} Same as
  Figure~\protect\ref{fig:biriinter600and585}, but for the BSM ratios
  $R_i^\mathrm{BSM}$, $i=2,\cdots,5$.}
\end{center}
\end{figure}

We used the quark propagators to calculate
\begin{eqnarray}
{\cal B}^1_{PP}(x_0,y_0)&=&
\frac{\sum_{\vec{x},\vec{y}}{\langle P(x) O_1(0) P(y)\rangle}}
{\frac83
\sum_{\vec{x},\vec{y}}{\langle P(x)\bar A_0(0)\rangle\langle 
\bar A_0(0)P(y)\rangle}}
\stackrel{a\ll x_0\ll T/2\ll y_0\ll T}{\longrightarrow} B_1 \\
{\cal B}^i_{PP}(x_0,y_0)&=&\frac{\sum_{\vec{x},\vec{y}}
{\langle P(x) O_i(0) P(y)\rangle}}
{N_i\sum_{\vec{x},\vec{y}}{\langle P(x)\bar P(0)\rangle\langle 
\bar P(0)P(y)\rangle}}
\stackrel{a\ll x_0\ll T/2\ll y_0\ll T}{\longrightarrow} B_i
\end{eqnarray}
for $i{=}2 \cdots 5$ with $N_i={\frac53,-\frac13,-2,-\frac23}$,
and fit to a constant in the symmetric time intervals given by: $12\le
x_0/a\le 19$ and $45\le y_0/a\le 52$ for $i{=}1 \cdots 5$, at
$\beta=6.0$; $10\le x_0/a\le 12$ and $36\le y_0/a\le 38$ for $i{=}1$
and $10\le x_0/a\le 14$ and $34\le y_0/a\le 38$ for $i{=}2\cdots 5$,
at $\beta=5.85$.

The bare values obtained for the parameters $B_i$ must be renormalized
to relate them to physical observables.  We used a non-perturbative
renormalization technique based on the RI/MOM methods of
\cite{Martinelli:1995ty}. Our results for the $B$-parameters are shown
in Figure~\ref{fig:biriinter600and585}, where the polynomial
interpolations to the kaon point are displayed. From these
$B$-parameters, we also reconstruct the matrix elements themselves. In
Figure~\ref{fig:riinter600and585}, we show the polynomial
interpolations to the kaon point of the ratios:
\begin{equation}
R_i^\mathrm{BSM}(\mu,M^2)
\equiv\left[\frac{F_K^2}{M_K^2}\right]_{expt}\left[\frac{M^2}{F^2}
\frac{\langle\bar P^0|O_i(\mu)|P^0\rangle}
{\langle\bar P^0|O_1(\mu)|P^0\rangle}\right]_{lat},
\label{eq:ridef}
\end{equation}
for $i=2,\cdots,5$, where $M$ and $F$ are the mass and ``decay
constant'' of the lattice kaon which we denote by $P^0$ to indicate
that the mass of the strange and down quarks that compose it can
differ from their physical values.  The ratios
$R_i^\mathrm{BSM}(\mu,M_K^2)$ measure directly the ratio of BSM to SM
matrix elements and, as such, can be used in expressions for $\Delta
M_K$ and $\epsilon$ beyond the SM, in which the SM contribution is
factored out.

Our main conclusion is that the non-SM, $\Delta S=2$ matrix elements
are significantly larger than found in the only other dedicated
lattice study of these amplitudes \cite{Donini:1999nn}. In tracing the
source of this difference, we found that we already disagree on the
much simpler matrix element of the pseudoscalar density between a kaon
state and the vacuum, which is the building block for the vacuum
saturation values of the BSM $\Delta S=2$ amplitudes. Through the
axial Ward identity, the matrix element of the pseudoscalar density is
related to the sum of the strange and down quark masses, which we find
to be roughly 30\% smaller than the value obtained in
\cite{Gimenez:1998uv}, with the same tree-level improved Wilson
fermion action and gauge configurations as used in
\cite{Donini:1999nn}. Since our result for this sum of masses is in
agreement with the continuum limit, benchmark result of
\cite{Garden:1999fg}, we are convinced that the stronger enhancement
of non-SM $\Delta S=2$ matrix elements that we observe is
correct. For details concerning this issue as well as our other
results, we refer the reader to \cite{Babich:2006bh}.

\section{Baryon wave-functions and diquark correlations}

We study the correlation of quarks inside baryons by evaluating baryon 
Green functions where the quarks, which without loss of generality 
we take to be $u, d$ and $s$, are found at different spatial locations 
at the sink:
\begin{equation}
G(\vec r_1, \vec r_2, \vec r_3, t)=
\langle u(\vec r_1, t) d(\vec r_2, t) s(\vec r_3, t)
\;\bar u(0)  \bar d(0) \bar s(0) \rangle 
\end{equation}
The color indices, which like the spin indices are left implicit,
are combined in a color singlet. The use of gauge fixing allows us 
to consider separated quarks without the need to introduce gauge 
transport factors.  For this investigation we also converted the gauge 
background field from the Landau gauge to the Coulomb gauge, and we will 
report below results in both gauges, which however are rather similar.  
In order to project over states of zero momentum, we sum over a 
translational degree of freedom, evaluating a reduced wave function
\begin{equation}
\tilde G(\vec r, \vec r', t)= \sum_{\vec r_3}
\langle u(\vec r_3 + \vec r, t) d(\vec r_3 + \vec r', t) s(\vec r_3, t)
\;\bar u(0)  \bar d(0) \bar s(0) \rangle 
\label{eq:gf}
\end{equation}
(The sums in \ref{eq:gf} involve a very large number of terms and the
use of a fast Fourier transform and the convolution theorem is crucial
to carry them out in manageable computer time). 

\begin{figure}[h!]
\begin{center}
\includegraphics*[width=0.7\textwidth]{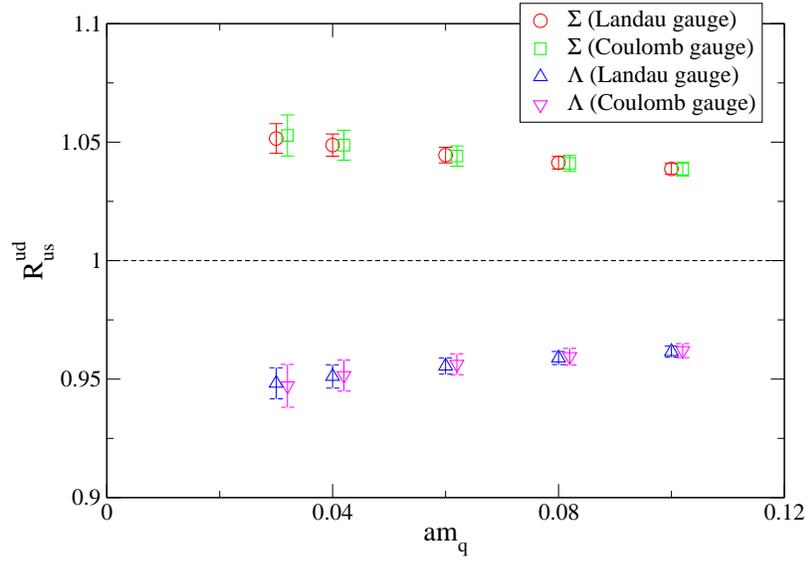}
\end{center}
\caption{\label{Rlight} Ratio of mean $u-d$ separation to mean $u-s$
  separation as function of quark mass for the two $u, d$ diquark
  configurations. Data with $\beta =6$ and $t=10a$.}
\end{figure}

The spin indices  are combined in an appropriate spin wave-function 
at the source and sink.  Of particular interest is comparing the spin 
configurations where, in the spin $1/2^+$ octet, the $u$ and $d$ quarks 
are combined, $\Lambda$-like , in a spin and isospin singlet state, 
the so-called good diquark combination, versus those where $u$ and $d$ 
are, $\Sigma$-like, in a spin and isospin triplet state.
In Figure~\ref{Rlight} we show the ratio of the mean separations
between $u, d$ and $u, s$ quarks in the two spin states. The results 
give support to the notion that quarks inside a baryon tend to 
correlate in a ``good'' diquark state. 

\begin{figure}[h!]
\begin{center}
\includegraphics*[width=0.4\textwidth,height=0.35\textwidth]{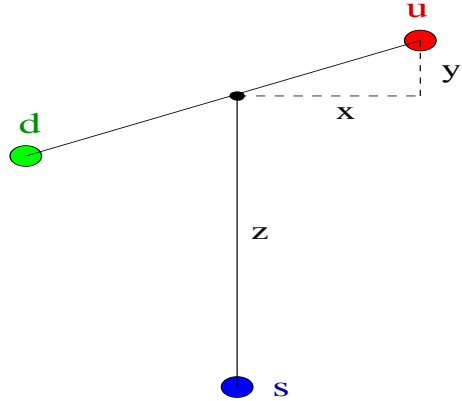}
\vspace{-5mm}
\end{center}
\caption{\label{geometry} The geometry used for visualizing the
  wave-function of the quarks inside a baryon.}
\end{figure}

\begin{figure}[h!]
\begin{center}
\includegraphics*[width=1\textwidth]{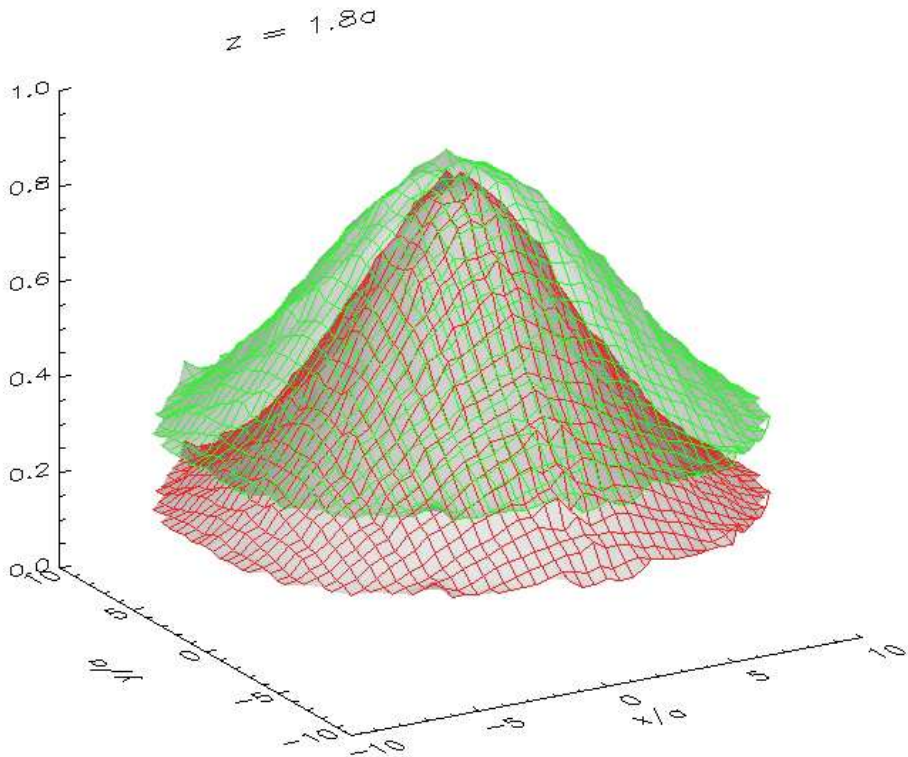}
\vspace{-5mm}
\end{center}
\caption{\label{fig-z4_whitebg} The wave-function of the $u$ and $d$
  quarks, in the Coulomb gauge, in a spin and isospin singlet state 
  (good diquark), given by the narrower bell-shaped surface (red with color),
  against the wave-function of the two quarks in a spin and isospin
  triplet state, given by the broader curve (green with color),
  for a separation $z=1.8a$ between the $s$ quark and the mid-point 
  of the $u, d$ pair.}
\end{figure}

\begin{figure}[h!]
\begin{center}
\includegraphics*[width=1\textwidth]{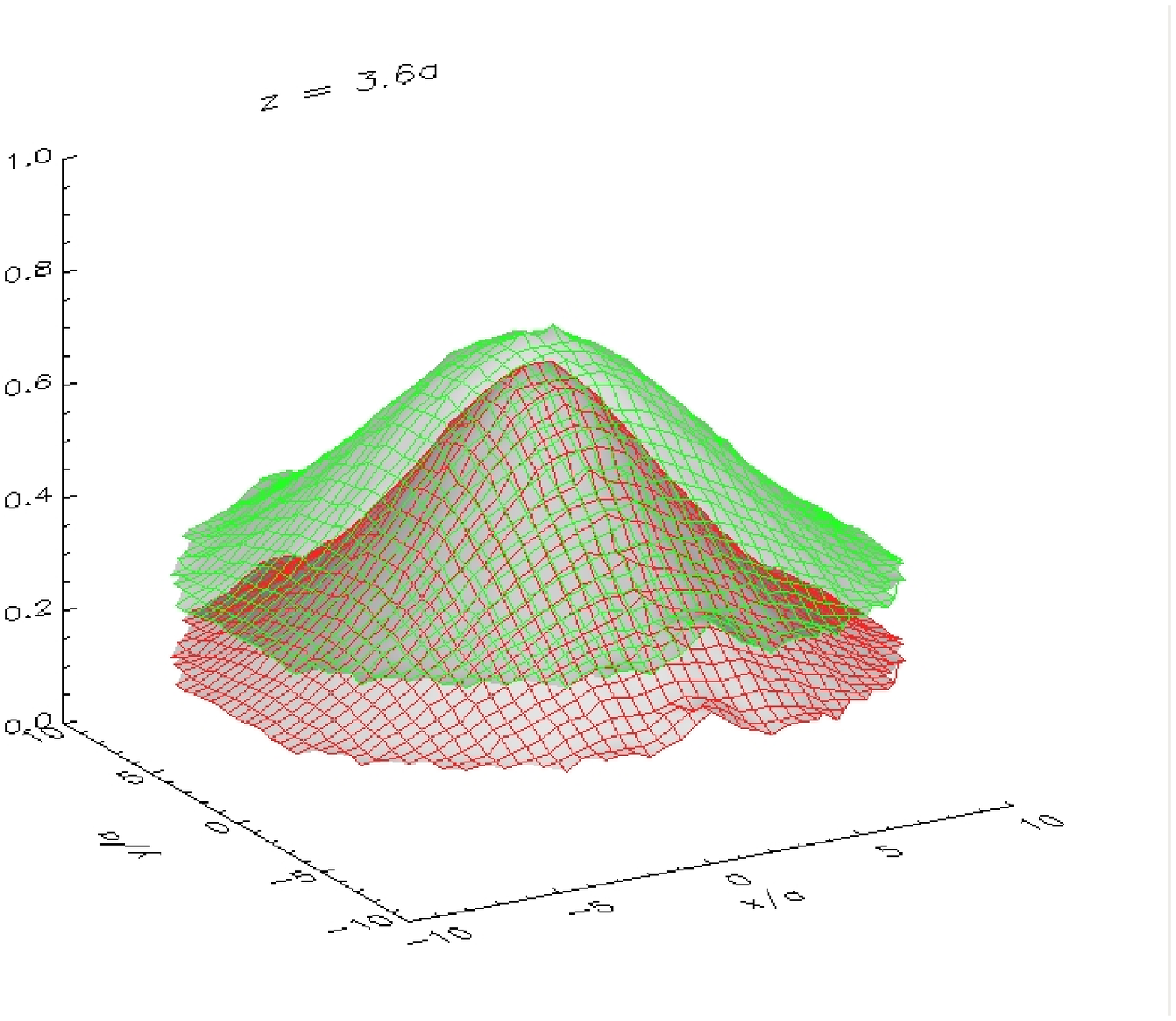}
\vspace{-5mm}
\end{center}
\caption{\label{fig-z8_whitebg} Same as in Figure 5, but for $z=3.6a$.}
\end{figure}
It is interesting to visualize the correlation among the quarks inside
the baryon, i.e.~the function $\tilde G(\vec r, \vec r', t)$ which
we can loosely consider as the wave-function of the three quarks.
The problem is, of course, that, at fixed $t$, $\tilde G$ is a 
function of the two vectors $\vec r, \vec r'$, i.e.~of six variables.
The representation of $\tilde G$ can be simplified somewhat by assuming
rotational invariance, i.e.~lack of major spin orbit correlation.
We have verified that this assumption is satisfied within the statistical 
accuracy of our calculations.  (Indeed the actual magnitude of  
spin-orbit correlations could be evaluated by our technique, given 
sufficient statistical precision.)  This leaves $\tilde G$ a function 
only of the shape of the triangle subtended by the two vectors 
$\vec r, \vec r'$.  This is still a function of three variables, 
but, to provide a meaningful visualization, we can fix one of 
these and represent the wave-function as a function of the other two.  
Thus, for a generic triangle formed by the locations of the 
$s, u$ and $d$ quarks, we introduce coordinates $x, y, z$ as
illustrated in Figure~\ref{geometry}, and then represent $\tilde G$
as a function in the $x, y$ plane at fixed $z$.  Our results, with
$\beta =6$, $a m_q=0.03$ and $t =10a$, are illustrated in 
Figures~\ref{fig-z4_whitebg} and \ref{fig-z8_whitebg}.  They show
again that the $u$ and $d$ quarks tend to correlate in the good 
diquark configuration.

A detailed, expanded version of the results presented in this section is
in preparation and will form the subject of a forthcoming paper.
An earlier study of quark wave-functions with some similarity to ours
can be found in Ref.~\cite{Hecht:1992ps}.  We are not aware of other
investigations studying the same type of correlation functions.

\section{Conclusions}

Our results corroborate the fact that the overlap discretization, 
at least insofar as valence quarks are concerned, can be used in large 
scale simulations, and, because of its very good symmetry properties, 
represents a choice method for QCD numerical calculations.

They also provide some novel matrix element values and evidence for 
strong diquark correlation in a flavor $\bar 3$, spin-singlet state.

\acknowledgments

This work was supported in part by US DOE grants DE-FG02-91ER40676, 
EU RTN contract HPRN-CT-2002-00311 (EURIDICE), and grant
HPMF-CT-2001-01468.  We thank Boston University and NCSA for use of their
supercomputer facilities.

\bibliographystyle{utcaps}
\bibliography{lat06_rebbi}

\end{document}